# Contract-based Hierarchical Resilience Management for Cyber-Physical Systems

Mohammad S. Haque^, Daniel Jun Xian Ng^, Arvind Easwaran^, Karthikeyan T.*

nusmsh@gmail.com, {danielngjj, arvinde}@ntu.edu.sg, karthik123@hotmail.com

^Nanyang Technological University and *Delta Electronics, Singapore

*ABSTRACT-* *Orchestrated collaborative effort of physical and cyber components to satisfy given requirements is the central concept behind Cyber-Physical Systems (CPS). To duly ensure the performance of components, a software-based resilience manager is a flexible choice to detect and recover from faults quickly. However, a single resilience manager, placed at the centre of the system to deal with every fault, suffers from decision-making overburden; and therefore, is out of the question for distributed large-scale CPS. On the other hand, prompt detection of failures and efficient recovery from them are challenging for decentralised resilience managers. In this regard, we present a novel resilience management framework that utilises the concept of management hierarchy. System design contracts play a key role in this framework for prompt fault-detection and recovery. Besides the details of the framework, an Industry 4.0 related test case is presented in this article to provide further insights.*

*KEYWORDS-* D.2.4.g Reliability < D.2.4 Software/Program Verification < D.2 Software Engineering < D Software/Software Engineering, D.2.15 Software and System Safety < D.2 Software Engineering < D Software/Software Engineering, D.2.4.f Programming by contract < D.2.4 Software/Program Verification < D.2 Software Engineering < D Software/Software Engineering

INTRODUCTION - Modern cyber-physical systems satisfy user requirements through a synergistic interaction between cyber and physical components. To maintain this synergy, the performance of components is not only important but also mandatory. However, large-scale CPS, such as Industry 4.0 based manufacturing plants, are massively distributed. In such systems, ensuring every component's functional and non-functional correctness is challenging.

When a cyber or physical component fails according to requirements, a fault is generated.

Depending on the source, faults can be broadly categorised into five categories. When system measurements are corrupted; causing a deviation between actual and sensed variables; the generated fault is a sensor fault [1]. Actuator fault represents discrepancy between the commanded and actual input to the system [1]. Process or component faults are failures of the system components, leading to changes in the system dynamics [1]. These include faults in both cyber as well as physical components and consider functional and non-functional requirements. When the communication channel among components fails to meet requirements (e.g., message deadlines), a communication fault is generated [2]. Environment fault occurs when something from the environment, but not a component of the CPS, causes a fault (e.g., particle strike).

A resilience management strategy is required to detect and recover from any fault. A primitive but widely used strategy is redundancy (e.g., hardware redundancy [3]). However, redundancy imposes considerable and unnecessary computational, spatial and energy burden [4]. Moreover, in massively distributed CPS, component redundancy introduces communication and synchronization overheads.

In contrast to redundancy, self-adaptive software-based resilience management approaches are more flexible and space-energy efficient [4-8]. Such approaches provide on-demand resilience services by assessing the status of the components and adapting them to meet requirements. However, some of these existing software-based approaches utilise a centralised architecture [4, 8]. In such architectures, components are connected to a central resilience management software (referred hereafter as "Resilience Manager" (RM)). Therefore, the RM's decision-making process slows down with an increase in the number of components. Moreover, a single RM has the risk of single-point failure. Due to this reason, in large-scale CPS, decentralised resilience management techniques are desirable.

A fully decentralised resilience framework requires fast collection of fault information and synchronization of decisions for quick detection and recovery. However, this is generally impractical in large-scale CPS, and hence existing decentralised approaches such as RIAPS [5] and iLand [6] attempt recovery without complete fault information. For example, the discovery service application in RIAPS does not consider the expected recovery time to a fault which incurs unnecessary communication overheads and delays [7]. Similarly, the reconfiguration manager in iLand is unaware of fault types and their recovery times which incurs unnecessary reconfigurations.

Thus, a resilience management framework that is scalable and detects and recovers from faults robustly is highly desirable. To this end, we propose and demonstrate hierarchical resilience management with contract-based fault detection framework in this paper. The hierarchy provides scalability, while formal contracts ensure robust and quick fault detection. Furthermore, RMs communicate information about the faults to enable more meaningful fault-recovery. We introduced the use of contracts for fault detection in [7], but that proposal neither includes a hierarchical framework nor does it consider the use of contracts with parameters; as discussed in the following sections these features are vital for scalability and efficient fault-recovery. Hierarchy allows decomposition of the resilience management functions, aiding scalability. Whereas, parametric contracts enable efficient runtime updates to the hierarchy so that system degradation techniques can be used for fault-recovery (e.g., reducing the speed of a conveyor belt when a robot is unable to pick up from the belt).

# Section 1. Hierarchical Resilience Management Framework

In this section, we present our new resilience management framework that utilises the concept of a management hierarchy. In this hierarchy, each manager makes use of parameterised system design contracts [9] to reduce communication overheads and execution time, while still maintaining scalability.

## Overview

In this framework, resilience management is the collaborative responsibility of a group of RMs. Each manager may have associated contracts that are used by observers to monitor the system for faults. A manager makes resilience decisions based on the information from its contracts as well as other managers. These managers follow a communication protocol that dictates to whom and what to talk about. Due to this protocol, a virtual management hierarchy of RMs and their associated contracts are created.

As shown in Figure 1, each RM in the leaf level is responsible for a cyber or physical component. It has observers which monitor whether the associated component satisfies its contracts. When a fault is detected in the component (contract failure), the manager first tries to recover locally. If no local solution exists, the manager informs its "parent RM." It may also provide fault information such as the extent of failure (e.g., amount of contract violation), and may receive updated parameter values for the contracts from its parent. If no parameter updates are received, it implies that the fault has been successfully handled at a higher-level in the hierarchy without any system degradation.

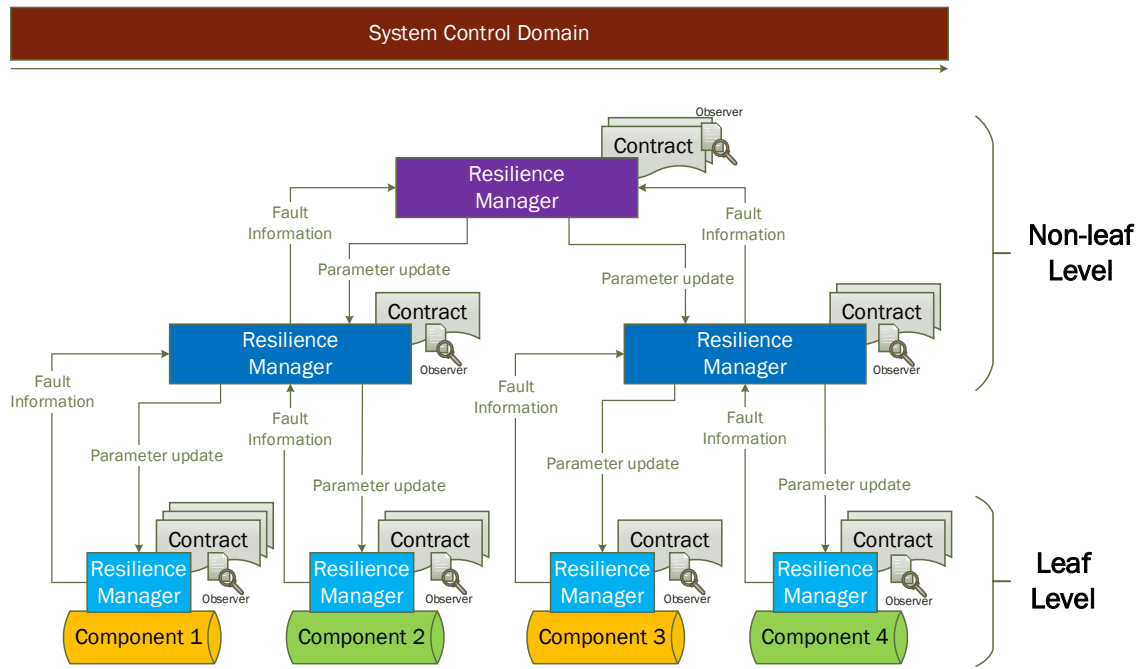

Figure 1 OVERVIEW OF THE FRAMEWORK

A non-leaf level RM, on the other hand, monitors whether multiple components duly collaborate to meet requirements. It receives fault information from lower-level managers and tries to recover locally. If the solution requires system degradation, new parameter values are generated and communicated to all the lower-level managers. If local recovery is not feasible, then fault information is further propagated upwards until the root level manager. If the root also fails to find a feasible solution, then it hands over responsibility to "system control," which is a decision logic outside the scope of the proposed framework (e.g., a human operator or resilient controller module).

**Framework Details**

The resilience framework is based on contracts for system design [9]. Such a contract is a precise description of (i) inputs and outputs of a component or subsystem, (ii) assumptions on the inputs and environment, and (iii) required guarantees about the outputs or behaviour of the component or subsystem. Our framework uses a variant of the system design contract, called "parametric contracts" [10], that allow parameterised specification of assumptions and guarantees so that these can be updated at runtime simply by changing the parameter values.

A set of contracts is assigned to a RM in our framework. For a given contract, runtime observers check whether the expected behaviour is generated by the component or subsystem of concern. Observers can be expressed using different formalisms like finite state machines and timed and hybrid automata [7]. Upon failure of a contract, a fault is triggered, and the manager performs an analysis to determine whether any solutions within its scope can be applied. This analysis could depend on the current number and extent of contract failures. If the chosen solution requires a parameter update, then the observer is informed accordingly. If no solution is found, the manager informs the higher-level RM, providing information about the fault. The higher-level manager, in turn, uses failure information from its own contracts as well as from lower-level managers to perform further fault-recovery. If the chosen recovery procedure at any level necessitates an update to contract parameters, the update is communicated to all the lower-level managers. Thus, throughout the process, only fault information and parameter values are exchanged between RMs.

User provided end-to-end requirements and component capabilities are used to generate the contracts. The component capabilities are first used to define leaf level contracts. For example, contracts for a controller component could be derived from information about the host hardware and available alternate controller behaviours. Contracts from different components in a subsystem could then be composed to derive higher-level contracts in the hierarchy. Contract parameters can be derived based on tuneable performance knobs available in the system. For example, in manufacturing plants, the speed of conveyor belts can be used as a contract parameter to degrade the plant's throughput. Functions based on these parameters could then be used in assumptions and guarantees of contracts. Although contract assumptions and guarantees could be defined using any desired logic, we restrict our focus to Boolean logic so that efficient observers can be implemented.

When contracts are composed, care must be taken to ensure that the resulting hierarchy satisfies desirable properties for contract composition and refinement (defined in [9] and reproduced below). *In particular, for this framework, we aim to ensure that the composition of a set of contracts belonging to some lower-level components/subsystems, is a refinement of the contract in the higher-level parent subsystem.* Additionally, it is also important to ensure that the root level contract satisfies (is a refinement of) user provided end-to-end requirements.

**Refinement of contracts:** A contract C' is a refinement of contract C when the following

conditions are satisfied:

1) Assumptions of C' are the weaker set of assumptions of C
2) Guarantees of C are the weaker set of guarantees of C'

**Composition of contracts:** Contracts $C_1$ and $C_2$ can be composed as $C_1 \otimes C_2$ when the following conditions are satisfied:

1) If the guarantees of one component ($C_{1/2}$) are independent of the assumptions of the other, then the assumptions of $C_1 \otimes C_2$ are the stronger of the assumptions of $C_1$ and $C_2$
2) If they are not independent, then the assumptions of $C_1 \otimes C_2$ are the weakest assumptions such that when they are conjuncted with the guarantees of $C_1$ (likewise $C_2$), the assumptions of $C_2$ (likewise $C_1$) are implied
3) Guarantees of $C_1 \otimes C_2$ are the conjunction of the guarantees of $C_1$ and $C_2$

Note, outputs of one component that are inputs of the other are omitted from $C_1 \otimes C_2$. Such a composition of contracts is useful when composing lower-level component contracts that form a cause-effect chain into a higher-level subsystem contract.

In Section 2, using a case study, we illustrate the process of contract generation and use them for providing resilience. Note, techniques for automatic generation of the contract hierarchy are out of the scope of this article and will be addressed in future work.

**Communication protocol between Resilience Managers:** Communication between a parent and child RM is similar to a client-server communication model. Client (child RM) initiates a request to a server (parent RM). The request contains fault information. The server responds by suggesting updates to contract parameters. In cases when the fault is handled without any performance degradation, the server may not provide any response. In cases when a "sibling client" issues a request, it is also possible to receive a response from the server without any self-initiated request.

**Advantages and Limitations of the Framework**

Only fault information and contract parameter values are exchanged by the RMs in our framework. Contract specifications do not change. Moreover, as soon as a solution is found within the hierarchy, the recovery process stops. As a result, communication overhead is reduced significantly when compared to existing fully centralised and decentralised architectures. Further, the hierarchical approach is also robust against single-point failures.

One of the challenges of the proposed framework is the design of the hierarchy. An efficient design would require information about component capabilities as well as how the components are combined to form subsystems. A key limitation is the inability to integrate stochastic approaches for fault detection. Both these issues will be addressed in future work.

# Section 2. Case Study

To evaluate the performance of our resilience framework, we implemented a testbed on a Fischertechnik Model (EAN-Code 4048962250404). The testbed can generate some fault scenarios observed in Industry 4.0 environment. In the following subsections, we present details about the testbed, the fault scenarios, the contract hierarchy monitored by our resilience framework and performance comparison of different resilience frameworks.

**Case Study Description**

Figure 2 presents a bird's eye view of the testbed. As shown, this model has two Light Sensors ($LS_1$, $LS_2$), a Colour Processor (CP), a conveyer belt, three ejectors and three bins to collect tokens. It also comes with a Motor Controller (MC) to control the speed of the motor that rotates the conveyor belt. Moreover, a Pulse Counter (PC) is present to count the number of steps that passed through $LS_1$.

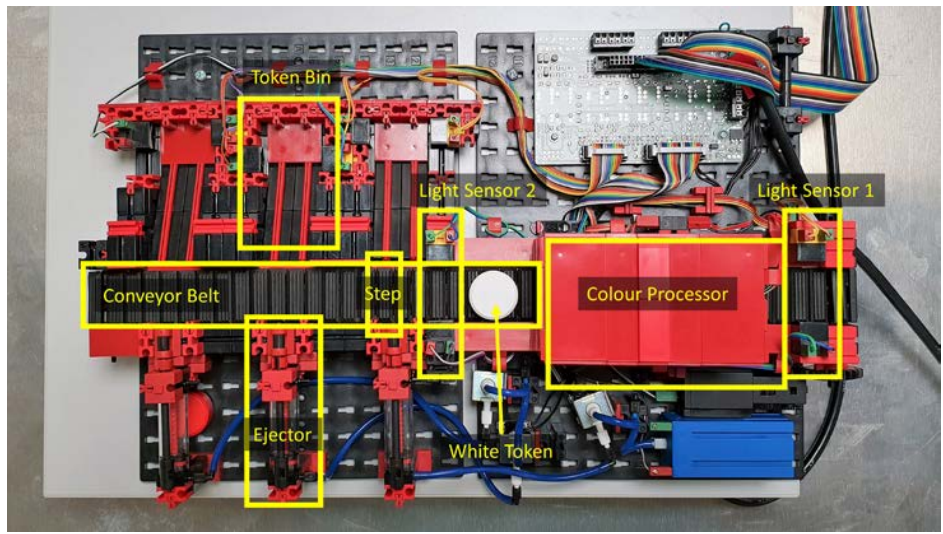

*Figure 2 FISCHERTECHNIK TRAINING MODEL (EAN-CODE 4048962250404)*

Figure 3 presents the operation flow of the training model. When a token is placed on the belt at $LS_1$, the sensor detects the presence of the token and activates CP. When the token passes through CP, the colour sensor detects the token's colour. A decision-making component; Bin Selector (BS), then calculates the step count at which the token reaches either the first, second or third ejector. In this calculation, the step number at which the CP was activated is used. The calculation result is sent to the Ejector Controller (EC). Moreover, depending on the colour of the token, the BS dictates the EC with regards to which ejector must be activated. The appropriate ejector activates as soon as the calculated step count is reached. *In our implementation, we only use the first 2 bins and white tokens. Under normal operation, the white token would be ejected into the first bin. In case of a fault that requires changes to the conveyor belt speed, the second bin is used temporarily until fault-recovery is completed.*

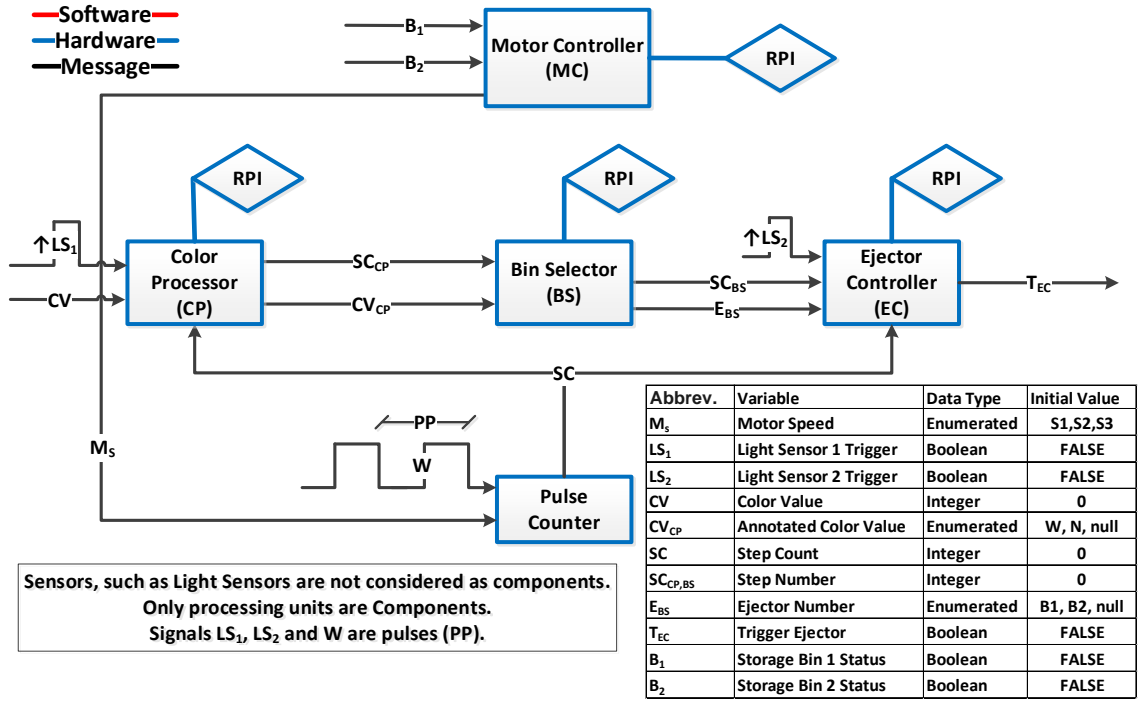

*Figure 3 FISCHERTECHNIK TRAINING MODEL REGULAR OPERATION FLOW*

**Logic Control Implementation**

As shown in Figure 3, instead of traditional PLCs, we introduced four Raspberry Pi 3 (RPI) to hold the control logic of CP, BS, EC, and MC. An RPI comes with 1.20GHz Quad Core Processor, 1GB RAM, and multiple General Purpose Input/Output pins. Each RPI runs Raspbian GNU/Linux 8.0 operating system (kernel version 4.9.35-v7). An Arduino Pro Mini microcontroller is also used to provide analogue to digital conversion of the colour sensor output for the RPI. In addition, due to the voltage differences between the RPIs and the testbed, voltage converters are used to interface them together. All the RPIs are interconnected over Ethernet through a network switch.

**Resilience Framework Implementation**

To introduce resilience management and fault injection capability on the testbed, we used FORTE, a runtime environment integrated with 4DIAC IDE [11], on the RPI. 4DIAC is based on the IEC 61499 standard, an event-driven function block model for distributed control systems. Our resilience framework, including observers for contracts and RMs, is built on top of 4DIAC.

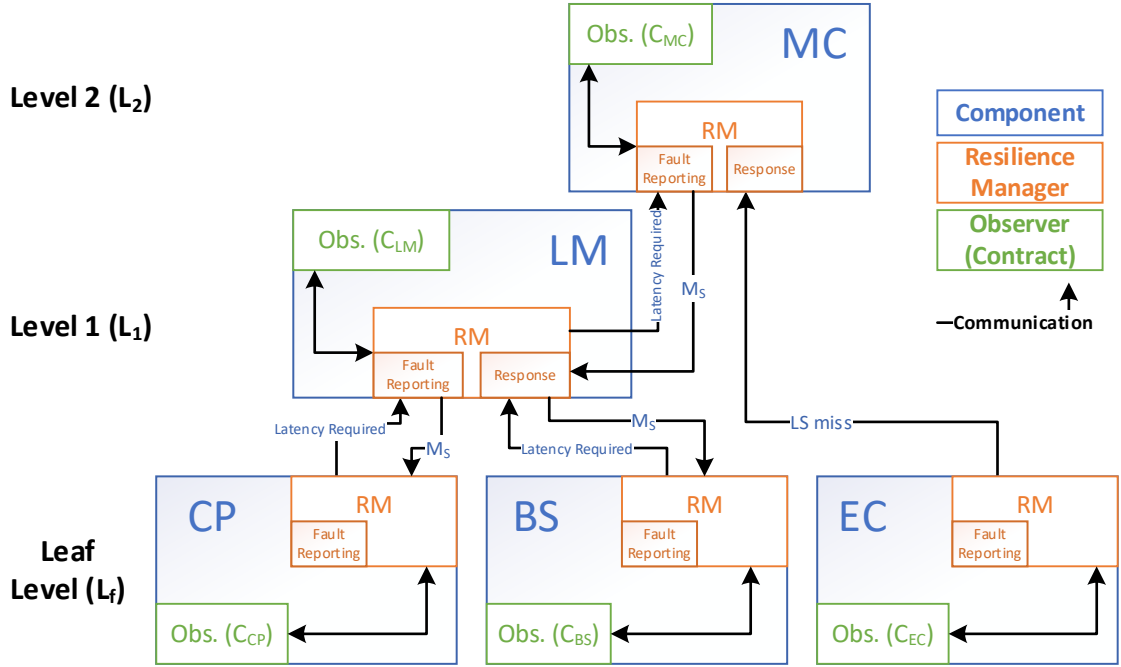

*Figure 4 ARCHITECTURAL DESIGN OF THE RESILIENCE MANAGEMENT FRAMEWORK*

Our resilience framework comprises of a three-level management hierarchy. There are three Leaf Level ($L_f$) RMs, one Level 1 ($L_1$) RM and one Level 2 ($L_2$) RM. The first $L_f$ uses a parameterised latency-related contract for CP and the second one uses a similar contract for BS (both correspond to a process fault scenario). The third $L_f$ on EC uses a contract that deals with the jitter of the token on the conveyor belt (corresponds to a physical fault scenario). $L_1$ handles the contract for Latency Management (LM) composed from the contracts of CP and BS. Finally, $L_2$ uses the root level contract composed of the contracts of $L_1$ and $L_f$ on EC. Leaf RMs reside on the RPIs corresponding to their respective components (CP, BS, and EC). $L_1$ resides on the RPI corresponding to BS, and $L_2$ resides on the RPI corresponding to MC because it controls the speed of the conveyor belt for fault-recovery.

Motor speed of the conveyor belt ($M_S$) is the only parameter used by all the contracts in the hierarchy except for the one on EC. This is reasonable because, given $M_S$ and the step count from PC, precise time durations for the token to reach specific sensors or components (e.g., $LS_2$, CP) can be computed.

**Contract and Resilience Manager Generation**

The contracts were manually generated based on user provided requirements and the capabilities of the components used. *The end goal of the user is to ensure that the tokens are correctly ejected in the bins.* For contract generation, we follow a bottom-up approach where we first identify the leaf components which would represent the lowest level of monitoring that can be done. Each component may have one or more contracts and corresponding observers associated with it. We then compose such leaf contracts into higher-level contracts to form the hierarchy. As described above, we identified three leaf contracts (corresponding to CP, BS, and EC), one $L_1$ contract which is composed of the contracts for CP and BS, and a $L_2$ contract which is composed of the contracts for $L_1$ and EC. Each contract except the one on EC uses $M_S$ as a parameter. We now present the contract and resilience hierarchy details. Please refer to Figure 3 for information on the variables used by these contracts.

A parametric contract named **C<sub>CP</sub>** is used by the $L_f$ RM of CP and is defined as follows.

*CONTRACT NAME: C<sub>CP</sub>*

- **INPUTS** --- $LS_1$
- **OUTPUTS** --- $SC_{CP}$; $CV_{CP}$
- **PARAMETERS** --- $M_S$
- **ASSUMPTIONS** --- $(M_S==S_1) \vee (M_S==S_2) \vee (M_S==S_3)$
- **GUARANTEE** --- $\uparrow LS_1 \Rightarrow (SC_{CP}!=0) \wedge (CV_{CP}!=null)$ within $f_{CP}(M_S)$

Contract **C<sub>CP</sub>** encodes the following information. It takes one input: light sensor trigger ($\uparrow LS_1$). For any allowed $M_S$ value (one of three), component CP should generate a step count ($SC_{CP}$) and an enumerated token colour value ($CV_{CP}$) within $f_{CP}(M_S)$ seconds. $SC_{CP}$ is the step number at which the CP was activated. When $CV_{CP}$ is 'W', it indicates a white token; 'N' indicates a non-white token and null means it has not been initialised.

Contract **C<sub>BS</sub>** is used by the $L_f$ RM of BS which takes two inputs: $SC_{CP}$ and $CV_{CP}$ generated by component CP. After receiving the inputs, component BS should generate two outputs within $f_{BS}(M_S)$ seconds: (i) ejector number to activate ($E_{BS}$) and (ii) step number at which the ejector should be activated ($SC_{BS}$). In our case th,is is the ejector for the first bin (normal operation).

*CONTRACT NAME: C<sub>BS</sub>*

- **INPUTS** --- $SC_{CP}$; $CV_{CP}$.
- **OUTPUTS** --- $E_{BS}$; $SC_{BS}$
- **PARAMETERS** --- $M_S$
- **ASSUMPTIONS** --- $(M_S==S_1) \vee (M_S==S_2) \vee (M_S==S_3)$
- **GUARANTEE** --- $(CV_{CP}!=null) \wedge (SC_{CP} != 0) \Rightarrow (SC_{BS}!=0) \wedge (E_{BS}!=null)$ within $f_{BS}(M_S)$

The observers for contracts **C<sub>CP</sub>** and **C<sub>BS</sub>** are monitoring the respective guarantees and also record the actual component execution latencies ($C_L$). The corresponding $L_f$ RMs have no fault-recovery mechanisms. When a contract fails, the RM directly reports the fault to $L_1$ RM together with information about the actual latency $C_L$ of the component.

The $L_1$ RM for Latency Management (LM) is responsible for the contract **C<sub>LM</sub>** defined below. It checks whether the guarantees of both contracts **C<sub>CP</sub>** and **C<sub>BS</sub>** are satisfied. This contract is generated using the contract composition technique described in Section 1. Further, to allow for dynamic slack management between CP and BS, contract refinement is used so that $f_{CP}(M_S) + f_{BS}(M_S) < f_{LM}(M_S)$. The difference between these two values is the latency slack available within $L_1$. If the reported actual latencies from lower-level RMs ($C_L$) are within this slack, then no further action is taken by $L_1$. Otherwise, it reports the fault to $L_2$ RM together with the amount of latency violation. This clearly shows the benefit of having contract hierarchy as latency can be dynamically distributed at runtime between CP and BS. Note, no observer is required for this contract because all the information is provided by lower-level RMs.

*CONTRACT NAME: C<sub>LM</sub> – Composition and refinement of C<sub>CP</sub> and C<sub>BS</sub>*

- **INPUTS** --- $LS_1$
- **OUTPUTS** --- $E_{BS}$; $SC_{BS}$
- **PARAMETERS** --- $M_S$
- **ASSUMPTIONS** --- $(M_S==S_1) \lor (M_S==S_2) \lor (M_S==S_3)$
- **GUARANTEE** --- $\uparrow LS_1 \Rightarrow (SC_{BS}!=0) \land (E_{BS}!=null)$ within $f_{LM}(M_S)$

A key design choice in the development of contracts **C<sub>CP</sub>**, **C<sub>BS</sub>** and **C<sub>LM</sub>** is the determination of various functions that map $M_S$ to latency values. In general, these functions for **C<sub>CP</sub>** and **C<sub>BS</sub>** should be such that they encode "typical" execution latencies for the components CP and BS, respectively. Whereas for **C<sub>LM</sub>**, the function should encode a time duration that is minimally the total typical execution latencies of CP and BS together, but no more than the time taken by the token to reach $LS_2$.

Contract **C<sub>EC</sub>** is used by the $L_f$ RM of EC, which checks for jitter of the tokens on the conveyor belt. It takes three inputs: (i) current step count SC, (ii) $SC_{CP}$ and (iii) the signal from $LS_2$. This contract checks if the rising edge of $LS_2$ coincides with the step count at $SC_{CP}$+*Offset*, where O*ffset* is the number of steps on the belt between CP and $LS_2$.

*CONTRACT NAME: C<sub>EC</sub>*

- **INPUTS** --- SC; $SC_{CP}$; $LS_2$.
- **OUTPUTS** --- None
- **PARAMETERS** --- None
- **ASSUMPTIONS** --- True
- **GUARANTEE** --- $\uparrow LS_2 \Leftrightarrow (SC==SC_{CP}+Offset)$

The observer for **C<sub>EC</sub>** monitors its guarantee and reports any failure to the $L_f$ RM of EC. This RM first reports the failure to $L_2$ and then activates the ejector for the second bin at the appropriate step count (temporary bin to collect tokens while fault-recovery is ongoing). The appropriate step count is determined once the $LS_2$ trigger is eventually received.

Finally, the root level contract **C<sub>MC</sub>** is used by the $L_2$ RM of MC. **C<sub>MC</sub>** is the composition of contracts **C<sub>LM</sub>** and **C<sub>EC</sub>**. The guarantee of this contract aims to ensure that all the tokens seen by $LS_1$ are successfully allocated a bin before the token reaches $LS_2$ and that the token indeed reaches $LS_2$ at the appropriate step count. Since $LS_2$ is the last sensor before the bins, this is the most reasonable monitoring that one could do to ensure user requirement satisfaction.

*CONTRACT NAME: C<sub>MC</sub>*

- **INPUTS** --- SC; $SC_{CP}$; $LS_1$; $LS_2$
- **OUTPUTS** --- $E_{BS}$; $SC_{BS}$
- **PARAMETERS** --- $M_S$
- **ASSUMPTIONS** --- $(M_S==S_1) \lor (M_S==S_2) \lor (M_S==S_3)$
- **GUARANTEE** --- $[\uparrow LS_1 \Rightarrow (SC_{BS}!=0) \land (E_{BS}!=null)$ within $f_{LM}(M_S)] \land$
  $[\uparrow LS_2 \Leftrightarrow (SC==SC_{CP}+Offset)$

$L_2$ finally combines information from all the contracts as well as lower-level RMs to decide whether

to reduce $M_S$ and hence degrade system performance. It first uses a bin sensor ($B_{1/2}$) to check whether the token was inserted into the correct bin ($E_{BS}$) irrespective of any reported faults. This may happen for instance when component BS is able to activate the ejector on time despite the latency violation of $L_1$ (there is still some time between the $LS_2$ trigger and the token reaching the first bin). Although unlikely, it may also happen when a positive jitter before the $LS_2$ trigger is accurately compensated by a negative jitter after the trigger and before the first bin. If not, and a fault is reported by $L_f$ of EC, then the motor speed is reduced to its slowest value of $S_3$ to eliminate this unpredictable jitter. Finally, if none of the above are true and a fault is reported by $L_1$, then the motor speed is reduced to either $S_2$ or $S_3$ depending on the amount of latency violation.

Thus, the hierarchical framework can progressively escalate faults until it eventually results in a system degradation. At each level, it attempts fault-recovery before escalation to minimise the impact on the system.

**Fault Scenarios**

Faults can be injected manually into the testbed through the RPIs and the 4DIAC function blocks of components to create the following scenario types:

1. For component CP, $C_L > f_{CP}(M_S)$, but $C_L + f_{BS}(M_S) \leq f_{LM}(M_S)$. That is, CP's $L_f$ RM reports a fault, but $L_1$ RM has sufficient slack.
2. Same as above for component BS.
3. Same as above for both components CP and BS. That is, both the $L_f$ RMs report faults, but $L_1$ RM still has sufficient slack.
4. $L_1$ RM reports a fault to $L_2$ RM.
5. EC's $L_f$ RM reports a fault to $L_2$ RM.
6. Both EC's $L_f$ RM and $L_1$ RM report faults to $L_2$ RM.

**Performance Evaluation**

Our experimental setup aims to compare the time and amount of communication required for fault-recovery in our resilience framework compared with representative (hypothetical) designs of fully centralised and fully decentralised resilience frameworks. Such frameworks do not have the concept of a management hierarchy.

In a fully centralised framework, any faults that occur in a component would be sent to a centralised manager which then responds accordingly. For the case study considered in this article, there would be three components (CP, BS, EC) and one centralised RM on the MC, requiring one message for fault reporting and three messages for a response (one to each component) for each fault.

A fully decentralised framework would have four components (CP, BS, EC, MC) with their RMs communicating with one another. A fault occurrence would require each RM to reach consensus before fault-recovery can happen. We assume the best design to reach consensus requires three message sets: (i) fault reporting, (ii) response with possible solutions and (iii) chosen solution. That is, nine messages in all for each fault in our test case. For both the centralised and decentralised frameworks, we assume that the components have the capability to detect local faults. This is to ensure that the comparison is fair.

We ran an experiment on our testbed that involved the injection of nine fault scenarios, where scenario types 1,2 and 3 occurred twice and types 4, 5 and 6 occurred once. This implies a total of 12 faults during the run because scenario types 3 and 6 generate two faults each. The number of messages generated by our framework during this run is 21. Whereas, for the fully centralised and decentralised frameworks this would result in 4*12=48 and 9*12=108 messages, respectively. This translates to communication savings of 56% and 81% when compared to the two architectures.

We now compare the time spent for fault-recovery. Measurements from our testbed indicate that the message latency is 1ms and decision-making is 0.5ms on an average. For the centralised and decentralised frameworks, each fault would require one and two decision-making steps respectively. For our framework, scenario types 1, 2, 3 and 5 require one decision-making step (either at $L_1$ or $L_2$), while types 4 and 6 require two steps (both at $L_1$ and $L_2$). Thus, the time spent in fault-recovery is 27.5ms, 54ms, and 114.5ms for our, the centralised and the decentralised frameworks, respectively.

# Conclusion

In this article, we present a first-of-its-kind resilience management framework that utilises the concept of management hierarchy and parametric contracts to reduce communication overhead and time for fault-recovery, while avoiding any single point failures. Experimental results reveal that this framework exchanges 56% and 81% fewer messages during fault-recovery compared to the fully centralised and decentralised architectures, respectively. Moreover, the hierarchical architecture makes it suitable for large-scale, Industry 4.0 like cyber-physical systems. Readers can watch the video in [12] to see how our resilience framework works for the fault scenarios discussed in this article.

# Acknowledgment

We would like to acknowledge the contributions from Sidharta Andalam, Delta Electronics Singapore, for his input on related works and formalisation of contracts. This research work was conducted within the Delta-NTU Corporate Lab for Cyber-Physical Systems with funding support from Delta Electronics Inc and the National Research Foundation (NRF) Singapore under the Corp Lab @ University Scheme.

## Author Bio

Mohammad Shihabul Haque was a Research Fellow in the School of Electrical and Electronic Engineering at Nanyang Technological University, Singapore, which he joined in 2016. He received his Ph.D. in Computer Engineering from the University of New South Wales, Australia, in 2008. His research interests lie in Real-time Memory Subsystems, Cyber-Physical Production Systems and Enterprise-level Hacking Techniques.

Daniel Jun Xian Ng is currently a Research Assistant in the School of Electrical and Electronic Engineering at Nanyang Technological University, Singapore, which he joined in 2016. He received a B.Eng. degree from Nanyang Technological University, Singapore, in Computer Engineering in 2016. His research interests lie in Automation, Cyber-Physical Systems and Embedded Systems.

Arvind Easwaran is currently an Associate Professor in the School of Computer Science and Engineering at Nanyang Technological University, Singapore, which he joined in 2013. He received a Ph.D. degree from the University of Pennsylvania, USA, in Computer and Information Science in 2008. His research interests lie in Cyber-Physical Systems, Real-Time Systems and Formal Methods. He is a member of IEEE and ACM.

Karthikeyan Thangamariappan is currently a Technologist in Rolls-Royce, Singapore. He received a MSc. degree from Nanyang Technological University, Singapore, in Communication Software and Networks (2009). His research interests lie in Closed Loop Control Systems and Predictive Maintenance.